\documentclass[aps,reprint,superscriptaddress,amsmath]{revtex4-1}

\usepackage{upgreek,graphicx}

\usepackage[normalem]{ulem}

\usepackage{xcolor,soul}

\newcommand{\ie}{i.e.}
\newcommand{\cf}{cf.}

\newcommand{\etal}{\textsl{et al.}}

\begin{document}

\title{Anisotropic diffusive transport: connecting microscopic scattering and macroscopic transport properties}

\author{Erik Alerstam}
\email{erik.alerstam@jpl.nasa.gov}
\affiliation{Division of Atomic Physics, Department of Physics, Lund University, \\
P.O. Box 118, 221 00 Lund, Sweden\\}
\affiliation{\textup{Author is currently at:} Jet Propulsion Laboratory, California Institute of Technology, \\ Pasadena, California 91109, USA\\ }
\date{\today}

\begin{abstract} 
This work concerns the modeling of radiative transfer in anisotropic turbid media using diffusion theory. A theory for the relationship between microscopic scattering properties (i.e., an arbitrary differential scattering cross-section) and the macroscopic diffusion tensor, in the limit of independent scatterers, is presented. The theory is accompanied by a numerical method capable of performing the calculations. In addition, a boundary condition appropriate for modeling systems with anisotropic radiance is derived. It is shown that anisotropic diffusion theory, when based on these developments, indeed can describe radiative transfer in anisotropic turbid media. More specifically, it is reported that solutions to the anisotropic diffusion equation are in excellent agreement with Monte Carlo simulations, both in steady-state and time-domain. This stands in contrast to previous work on the topic, where inadequate boundary conditions and/or incorrect relations between microscopic scattering properties and the diffusion tensor have caused disagreement between simulations and diffusion theory. The present work thus falsify previous claims that anisotropic diffusion theory cannot describe anisotropic radiative transfer, and instead open for accurate quantitative diffusion-based modeling of anisotropic turbid materials.
\end{abstract}

\maketitle

\section{Introduction}
In contrast to radiative transfer in general, the study of anisotropic diffusive transport of light has a relatively short history. Research into this subfield of radiative transfer initially began with the research on light transport in nematic liquid crystals some 20 years ago  \cite{Kao1996_PRL, Stark1996_PRL, VanTiggelen1996_PRL}. Anisotropic diffusive transport has since then been a topic of active research, and the phenomenon has been observed in a wide range of random media, for example, porous semiconductors \cite{Johnson2002_PRL}, stretched plastics \cite{Johnson2008_OptExpress, Johnson2009_JBO}, wood \cite{Tsuchikawa2002_ApplSpectrosc,Kienle2008_OptExpr}, various biological tissues \cite{Nickell2000_PhysMedBiol, Sviridov2005_JBO, Kienle2006_PRL}, and recently in compacted granular materials \cite{Alerstam2012_PRE}. 
As with multiple scattering phenomena in general, diffusive transport in anisotropic random media may be modeled with radiative transport theory (RTT). In RTT, the transport is described by the radiative transport equation (RTE) along with the microscopic scattering properties, which are directly linked to material microstructure. Considering transport on a macroscopic scale, most of the details of the microscopic scattering are averaged out by multiple scattering. Thus, when modeling macroscopic systems, for example, for comparison with experimental data, the RTE is often reduced to an (anisotropic) diffusion equation (e.g., Ref. \cite{vanTiggelen2000_RevModPhys}). Here, the macroscopic transport is characterized solely by the sample geometry and a diffusion tensor, $\mathbf{D}$, which embodies the sample microscopic scattering properties and provides a physical link between the microscopic (RTE) and the macroscopic (diffusion) models. However, in contrast to isotropic theory, where a simple relationship between the microscopic scattering properties and macroscopic properties (i.e., the diffusion coefficient) exist, this relationship is not yet firmly established for anisotropic systems. A notable exception is the special case of nematic liquid crystals, which has been extensively theoretically investigated, see, e.g., the review by Van Tiggelen and Stark  \cite{vanTiggelen2000_RevModPhys}.

Despite the importance of the various materials exhibiting transport anisotropy, and consequently, the necessity for accurate physical models of anisotropic transport, surprisingly little work has been performed to investigate the connection between the microscopic- (RTT) and macroscopic- (diffusion) models: Heino \etal\ compared solutions to the RTE (solved using the Monte Carlo method) and the anisotropic diffusion equation (in the frequency domain), observing a reasonable agreement \cite{Heino2003_PRE}. In a more extensive study, Kienle observed significant deviations between Monte Carlo simulation results and anisotropic diffusion theory when making comparisons in the steady-state and time domain. As a result, he argued that the anisotropic diffusion cannot be used to model anisotropic radiative transport \cite{Kienle2007_PRL}. It was pointed out by Johnson and Lagendijk that this conclusion, and the lack of agreement in the comparison, was due to an incorrect relationship between the scattering cross-section and the diffusion tensor  \cite{Johnson2009_JBO}. Furthermore, they provided an analytical method for calculating a diffusion tensor from a differential scattering cross section and compared anisotropic diffusion theory with experimental steady-state measurements from well-characterized samples (fibrous and stretched plastics). However, no comparison between direct solutions to the RTE and the anisotropic diffusion equation was made. Instead, the validation was limited to two comparisons with experimental data, out of which only one did show good agreement \cite{Johnson2009_JBO}. Recently, Kienle \etal\ revisited the subject, reporting on an extensive comparison between Monte Carlo solutions of the RTE and anisotropic diffusion theory. Even when designating the diffusion tensor elements as free parameters when fitting anisotropic diffusion theory solutions to Monte Carlo simulation results (thus effectively circumventing the need to know the exact relationship between the differential scattering cross section and the diffusion tensor), significant disagreements were still observed both in the steady-state and time domain \cite{Kienle2013_PhysMedBiol}. In conclusion, the current literature suggests that anisotropic diffusion theory cannot be used to accurately describe anisotropic radiative transport.

In this paper, careful treatment of the relation between microscopic scattering properties and macroscopic transport shows that anisotropic diffusion theory indeed is capable of accurately describe anisotropic radiative transfer. This result stands in contrast to the findings of Kienle \etal, and significant efforts are therefore devoted to elucidating shortcomings of earlier treatments of anisotropic diffusion. First, it is shown that macroscopic anisotropic transport is associated with an anisotropic radiance that invalidates results borrowed from isotropic diffusion theory. In particular, it is shown that the simplistic Òisotropic-typeÓ relation between the microscopic scattering properties and the macroscopic diffusion tensor does not hold. Moreover, an anisotropic radiance invalidates standard boundary conditions, even in the absence of a refractive index mismatch. A theory, founded in intuitive random walk theory, is presented as a way to connect the microscopic structure (i.e., an arbitrary differential cross section) with the diffusion tensor, in the limit of independent scatterers. Furthermore, the commonly used extrapolated boundary condition is modified to account for the anisotropic radiance. Using these new methods for calculating the diffusion tensor and the boundary condition, anisotropic diffusion theory is compared to results from Monte Carlo simulations (both in steady-state and time domain) for a wide variety of isotropic and anisotropic systems. As already implied above, the agreement between anisotropic diffusion theory and radiative transfer simulations is found to be excellent.

The remainder of the paper is organized as follows. Section \ref{sec:anisoRadiance} first provides a summary of microscopic scattering properties (the differential cross-section) (\ref{subsec:diffscattcrosssec}) and elaborate how these relate to the anisotropic radiance (\ref{subsec:aniso_radiance}). Section \ref{sec:MC} presents the model for how anisotropic systems are implemented in this work and describes how Monte Carlo simulations of anisotropic transport have been carried out. Section  \ref{sec:diff} starts with a short review of anisotropic diffusion theory (\ref{subsec:anisoDE}) and then moves on to one of the central parts of this work Ð how the diffusion tensor should be calculated from arbitrary microscopic scattering properties (\ref{sec:diffusion_tensor}) and the derivation of the extrapolation length for anisotropic turbid media (\ref{sec:ze}). Section \ref{subsec:Numest} discusses how, in the absence of analytical solutions to the presented theory, the diffusion tensor and extrapolation length can be estimated numerically. Section \ref{sec:Dmusg} elucidates the relation between the microscopic scattering properties and the (macroscopic) diffusion tensor, and Sec. \ref{sec:diffvsMC} reports on the comparisons between Monte Carlo simulations and anisotropic diffusion theory. Finally, Sec. \ref{sec:discussion} provides a general discussion of the results.

\section{Anisotropic radiance}\label{sec:anisoRadiance}
This section gives a brief summary of microscopic scattering properties and how they relate to the radiance. For a more detailed discussion on the topic, the reader is directed to, for example, the book by Welsh and van Gemert \cite{Welch2010_Book}.
\subsection{The differential scattering cross section}\label{subsec:diffscattcrosssec}
In this work, it is assumed that transport in the considered systems can be accurately modeled using radiative transport theory (RTT). Here it is assumed that the system can be reduced to a collection of discrete, independent, scatterers, each characterized by a differential scattering cross section, $\partial\sigma_\mathrm{sc}(\mathbf{\hat{s}'},\mathbf{\hat{s}})/\partial\Omega$, which describes the energy transferred in the direction of $\mathbf{\hat{s}'}$ when a plane wave, traveling along $\mathbf{\hat{s}}$, interacts with the scatterer. It may be calculated analytically in simple cases (spheres, cylinders, etc.) or computed numerically for arbitrary scatterers, using, for example FDTD. 

The differential scattering cross section is used to calculate two important physical quantities; the scattering coefficient,
\begin{equation}\label{eq:mus}
\mu_\mathrm{s}(\mathbf{\hat{s}}) = \sigma_\mathrm{sc}(\mathbf{\hat{s}})\rho_\mathrm{sc},
\end{equation}
and the single-scattering phase function,
\begin{equation}\label{eq:p}
p(\mathbf{\hat{s}'},\mathbf{\hat{s}}) = \frac{\partial\sigma_\mathrm{sc}(\mathbf{\hat{s}'},\mathbf{\hat{s}})}{\partial\Omega}\frac{1}{\sigma_\mathrm{sc}(\mathbf{\hat{s}})}.
\end{equation}
Here $\rho_\mathrm{sc}$ is the density of scatterers and $\sigma_\mathrm{sc}(\mathbf{\hat{s}})$ is the scattering cross section:
\begin{equation}\label{eq:sigma}
\sigma_\mathrm{sc}(\mathbf{\hat{s}}) = \int\limits_{4\pi}\frac{\partial\sigma_\mathrm{sc}(\mathbf{\hat{s}'},\mathbf{\hat{s}})}{\partial\Omega}\mathrm{d}\Omega'.
\end{equation}
It is easy to realize that if either the scattering coefficient or the single-scattering phase function is dependent on the current direction, $\mathbf{\hat{s}}$, the resulting macroscopic transport will be anisotropic. Furthermore,  for any physically realistic system (whose scattering parameters, $\mu_\mathrm{s}(\mathbf{\hat{s}})$ and $p(\mathbf{\hat{s}'},\mathbf{\hat{s}})$, are derived from the same differential scattering cross section) it is likely that if the scattering coefficient is direction-dependent, then so is the single-scattering phase function, and vice versa. Thus, any theory that connects the differential scattering cross section with the diffusion tensor must take both of these physical properties into account. 

\subsection{Anisotropic radiance} \label{subsec:aniso_radiance}
The radiance, $L(\mathbf{r},\mathbf{\hat{s}})$ [W/m$^2$sr], is a fundamental prosingle-scatteringperty in radiative transport.
It describes the power per unit area per solid angle traveling in direction $\mathbf{\hat{s}}$ at position $\mathbf{r}$, and is the product of the energy density $W(\mathbf{r})$ [J/m$^3$], the energy transport velocity, $v(\mathbf{\hat{s}})$ [m/s], in direction $\mathbf{\hat{s}}$, and a function $P(\mathbf{\hat{s}})$ [1/sr] describing the angular distribution of the radiance:
\begin{equation}\label{eq:radiance}
L(\mathbf{r},\mathbf{\hat{s}})=W(\mathbf{r})P(\mathbf{\hat{s}})v(\mathbf{\hat{s}}).
\end{equation} 
Here it was assumed that the angular distribution of the radiance is independent of the position, $\mathbf{r}$. The decomposition of the radiance reveals the two different means by which macroscopic anisotropic transport may arise: by a direction-dependent energy transport velocity (caused by, for example, an anisotropic refractive index) or through an anisotropic angular distribution, $P(\mathbf{\hat{s}})$. Intuitively (adopting the random walk picture of diffusive transport),  $P(\mathbf{\hat{s}})$, may be interpreted as a probability density function describing the probability that a random walker is traveling in direction $\mathbf{\hat{s}}$. It is clear that anisotropy in $P(\mathbf{\hat{s}})$ is a direct consequence of a direction-dependent scattering coefficient and/or single-scattering phase function: If the anisotropy is caused by a direction-dependent single-scattering phase function, some directions of travel will be more probable than others while all random walkers will, on average, take steps of the same length regardless of direction. If, on the other hand, the anisotropy is caused by a direction-dependent scattering coefficient, all directions will be equally represented on a step-by-step basis, but random walkers will take longer steps (spending more time to complete the step) in some directions, causing $P(\mathbf{\hat{s}})$ to be anisotropic.\\

\begin{figure}
\begin{center}
  \includegraphics{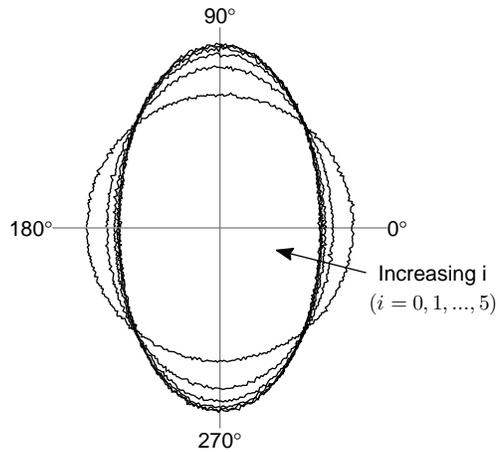}
 \caption{A polar plot (arbitrary units) of $P(\phi)$, the azimuthal component of  $P(\mathbf{\hat{s}})=P(\theta)P(\phi)$, as a function of number of steps, $i$.  In this system $P(\theta)$ uniformly distributed  $\in[0,\pi]$ and is thus not shown. The anisotropy is caused by a direction-dependent single-scattering phase function (with the tensor elements $g_{xx}=0.4$, and $g_{yy}=g_{zz}=0.8$, as defined in Sec. \ref{sec:MC}), while the scattering coefficient is direction independent. Starting from an isotropic distribution (at $i=0$), it takes many steps for the distribution to converge to the stationary distribution (full convergence not shown). }\label{fig:Radiance}
\end{center}
\end{figure}

The angular distribution of the radiance, $P(\mathbf{\hat{s}})$, is easily calculated in the simple case where only the scattering coefficient is direction-dependent. However, as discussed in the previous section, this is unlikely in any realistic physical system. When the single-scattering phase function is direction-dependent, the calculation of $P(\mathbf{\hat{s}})$ is no longer trivial. In fact, the stationary angular distribution is given by the solution to a Markov chain problem with a continuous state space, where the single-scattering phase function gives the probabilities for scattering from and to all directions (states) and the (direction-dependent) scattering coefficient describes the weights for each direction (how long the system will stay in a certain state). Fortunately, the Monte Carlo method gives a simple and straightforward solution to this difficult problem. The state space is seeded with an arbitrary source function and propagated step by step according to the probabilities described by $\mu_\mathrm{s}(\mathbf{\hat{s}})$ and $p(\mathbf{\hat{s}'},\mathbf{\hat{s}})$ until a stationary solution is reached. The result of such a numerical calculation is illustrated in Fig. \ref{fig:Radiance}. In this example, for illustration purposes, the angular radiance distribution, $P(\mathbf{\hat{s}})$, is converted to spherical coordinates and decomposed into two functions, $P(\mathbf{\hat{s}})=P(\theta)P(\phi)$, dependent on the polar angle, $\theta$, with respect to the $z$ axis, and the azimuthal angle, $\phi$. The initial distribution (at step $i=0$) is an isotropic distribution. For each step the distribution approaches the anisotropic stationary distribution. The consequence of the resulting anisotropic radiance, and the non-instantaneous convergence to a stationary distribution will be investigated in Sec. \ref{sec:diff}.

\section{Monte Carlo simulations of anisotropic transport}\label{sec:MC}
In this work, the Monte Carlo method, which provides a direct solution to the RTE, serves as the gold standard model to facilitate a comparison between RTT and anisotropic diffusion theory. A standard GPU-accelerated Monte Carlo code \cite{Alerstam2008b_JBO, Alerstam2010_BiomedOptExpress} was modified to accommodate anisotropic transport. The code allows for two different types of anisotropy, a direction-dependent scattering coefficient, $\mu_s(\mathbf{\hat{s}})$, and a direction-dependent single-scattering phase function, $p(\mathbf{\hat{s}'},\mathbf{\hat{s}})$, where $\mathbf{\hat{s}}= [dx\ dy\ dz]$ and $\mathbf{\hat{s}'}$ are a unit vectors describing the current and outgoing direction, respectively.

In the code the direction-dependent scattering coefficient is implemented as $\mu_s(\mathbf{\hat{s}}) = \mathbf{\hat{s}}\boldsymbol{\mu}_s\mathbf{\hat{s}}^{\mathrm{T}}$, where
\begin{equation}\label{eq:mu_tensor}
\boldsymbol{\mu}_s = \left( \begin{array}{ccc}
\mu_{s,xx} & 0 & 0 \\
0 & \mu_{s,yy} & 0 \\
0 & 0 & \mu_{s,zz} \end{array} \right),
\end{equation}
is the scattering coefficient tensor. The direction-dependent scattering phase function is implemented using the well known Henyey-Greenstein scattering phase function \cite{Henyey1941_AstrophysJ}, while allowing a direction-dependent scattering anisotropy factor $g(\mathbf{\hat{s}}) = \mathbf{\hat{s}}\mathbf{g}\mathbf{\hat{s}}^{\mathrm{T}}$ with
\begin{equation}\label{eq:g_tensor}
\mathbf{g} = \left( \begin{array}{ccc}
g_{xx} & 0 & 0 \\
0 & g_{yy} & 0 \\
0 & 0 & g_{zz} \end{array} \right).
\end{equation}
It is important to note that this implementation does not necessarily represent a physically realistic case, where the scattering properties are calculated from the same differential scattering cross section [Eqs. \ref{eq:mus}-\ref{eq:sigma}]. Rather, it serves as a test case where the two different means to create an anisotropic radiance are separately tunable. Also, this test case provides well-characterized results when the system is isotropic (\ie, $\mathbf{g}$ and $\boldsymbol{\mu}$ are scalar matrices) and makes the results comparable to those of previous studies. For clarity and brevity of the following discussion, and for easily interpreted results, $\boldsymbol{\mu}_s$ and $\mathbf{g}$ are restricted to being diagonal. Also, for simplicity, the refractive index of the medium, $n$, is assumed to be isotropic in this work. Neither of these assumptions imply restrictions in the theory. 

The code records the spatial and temporal characteristics of light transmitted through a slab (of thickness $L$ in the $z$ direction), infinitely extended in the $xy$ directions. At time $t=0$, light (unpolarized) is injected at $(x,y,z)=(0,0,0)$ and directed in the $z$-direction. The code uses Fresnel's formula (averaged over polarizations) for handling boundary transitions from the random media (with refractive index $n$) to the surrounding media $n_0=1.0$. Owing the temporally resolved detection, absorption can be added post-simulation using Beer-Lambert's law, but in this work $\mu_a=0$ for simplicity.
 \vspace{1 mm} 

\section{Theory for anisotropic diffusion}\label{sec:diff}
\subsection{The anisotropic diffusion equation}\label{subsec:anisoDE}
The three-dimensional time-dependent anisotropic diffusion equation in the absence of absorption reads \cite{vanTiggelen2000_RevModPhys,Wiersma1999_PRL, Alerstam2011_PhDThesis}
\begin{equation}\label{eq:AnisoDiff}
\frac{\partial W(\mathbf{r},t)}{\partial t} = \nabla\cdot\mathbf{D}\nabla W(\mathbf{r},t) + S(\mathbf{r},t),
\end{equation}
where $\mathbf{r}$ is the position, $t$ is time, $W(\mathbf{r},t)$ is the energy density, and $S(\mathbf{r},t)$ is a source function. Note that this equation may be derived without assuming a near-isotropic radiance, as commonly done when deriving the isotropic diffusion equation. $\mathbf{D}$ is the anisotropic diffusion tensor, which, in this work is constrained to the case where the matrix is diagonal:
\begin{equation}
\mathbf{D} = \left( \begin{array}{ccc}
D_{xx} & 0 & 0 \\
0 & D_{yy} & 0 \\
0 & 0 & D_{zz} \end{array} \right).
\end{equation}\\
This implies that the principal axes of the anisotropic radiance ellipsoid (\cf\ Fig. \ref{fig:Radiance}) are aligned with the axes of the cartesian coordinate system. Note that is not a limitation of the theory as diffusion tensors are diagonalizable by rotation \cite{Johnson2009_JBO, Dudko2005_BioPhysJ}.\\

In order to compare the results of the previously described Monte Carlo code, and anisotropic diffusion theory, Eq. (\ref{eq:AnisoDiff}) is solved for transmission through a slab of thickness $L$, infinitely extended in the $xy$-direction. Boundaries are handled by imposing the extrapolated boundary condition \cite{Keijzer1988_ApplOpt,Zhu1991_PRA}: $W=0$ at $z=-z_e$ and $z = L+z_e$, and using the source function $S(\mathbf{r},t) = \delta(x)\delta(y)\delta(z-z_0)\delta(t)$. The distance $z_0$ is the source depth, approximated by the transport mean free path in the $z$ direction, $z_0={\ell_z}^{\ast}$. In turn, ${\ell_z}^{\ast}$ is related to the diffusion tensor through
\begin{equation}\label{eq:Dzzell}
D_{zz} = \frac{1}{3}v_z{\ell_z}^{\ast},
\end{equation}
where $v_z$ is the energy transport velocity in the $z$ direction (in the absence of resonant scattering assumed to be $v_z = c/n_z$, where $n_z$ is the refractive index in the $z$-direction). The distance $z_e$ is called the extrapolation length and is usually derived assuming a near-isotropic radiance. As shown later (Sec. \ref{sec:ze}), the extrapolated boundary condition is still applicable in anisotropic diffuse systems, as long as proper care is taken when calculating $z_e$. The time evolution of the intensity transmitted through the slab reads \cite{Wiersma1999_PRL,Wiersma2000_PRE}:
\begin{widetext}
\begin{align}\label{eq:I_tr}
&I_\textrm{tr}(x,y,t)= \frac{I_0 \exp\left(-\frac{x^2}{4D_{xx}t}\right)\exp\left(-\frac{y^2}{4D_{yy}t}\right)}{\pi^{3/2}(4t)^{5/2}\sqrt{D_{xx}D_{yy}D_{zz}}} \times\left[ \sum_{m=-\infty}^{+\infty} z_{+,m} \exp\left(\frac{-z_{+,m}^2}{4D_{zz}t}\right) -z_{-,m} \exp\left(\frac{-z_{-,m}^2}{4D_{zz}t}\right)\right], \\
& \textrm{with} \begin{cases}
   z_{+,m}=L(1-2m) -4mz_e-z_0 \\
    z_{-,m}=L(1-2m) -(4m-2)z_e+z_0
 \end{cases}\nonumber
\end{align}
\end{widetext}

\subsection{Deriving the diffusion tensor}\label{sec:diffusion_tensor}
The relationship between the macroscopic diffusion tensor, $\mathbf{D}$, and the microscopic differential scattering cross section (\ie, the microscopic scattering properties $\mu_s(\mathbf{\hat{s}})$ and $p(\mathbf{\hat{s}'},\mathbf{\hat{s}})$, or in this case, $\boldsymbol{\mu}_s$ and $\mathbf{g}$), is the subject of confusion in the literature \cite{Kienle2007_PRL, Kienle2013_PhysMedBiol, Johnson2009_JBO}. Part of this confusion stems from the simple relationship between the two parameters in the case of isotropic diffusion:
\begin{equation}\label{eq:Diso}
D^\mathrm{iso} = \frac{c}{3n\mu_s(1-g)}=\frac{c}{3n\mu_s'},
\end{equation}
where $g$ is the first moment of the probability density function describing the scattering deflection angle (\ie, using a direction independent single-scattering phase function). Failing to note that this relationship is only valid for isotropic systems has lead to the adoption of the same (simplistic) relationship when considering anisotropic systems:
\begin{equation}\label{eq:Dsimplistic}
D_{kk}^{\mathrm{simplistic}} = \frac{c}{3n\mu_{s,kk}(1-g_{kk})} = \frac{c}{3n\mu_{s,kk}'}.
\end{equation}
In reality, the relationship between the macroscopic transport properties and the microscopic scattering properties is more complicated. This section outlines how random walk theory can be used to intuitively connect an arbitrary transport problem (defined by the differential scattering cross section, and the direction-dependent energy velocity) to a diffusion tensor. Further details on this method are presented in Ref. \cite{Alerstam2011_PhDThesis}, which in turn is influenced by the tutorial by Vlahos \etal\ \cite{Vlahos2008_OrderandChaos}. For brevity and clarity, it is assumed that the axes of the radiance ellipsoid (\cf\ Fig. \ref{fig:Radiance}) are aligned with the cartesian coordinate system; \ie, $\mathbf{D}$ is diagonal.\\

Consider a random walker in an unbounded three-dimensional absorptionless medium. The random walk is characterized by the scattering coefficient, phase scattering function, refractive index, etc., as in a normal Monte Carlo simulation. To determine the diffusion tensor element $D_{xx}$, consider the position of the random walker projected onto the $x$ axis. After $N$ steps the position is:
\begin{equation}\label{eq:Diffusion:X-sum}
x_N = \Delta x_1 + \Delta x_2 + \Delta x_3 + ... + \Delta x_N = \sum_{k=1}^{N} \Delta x_k. 
\end{equation}
The increments $\Delta x_i$ are random variables (dependent, unless the scattering is isotropic), each representing the projection of step $i$ onto the $x$ axis.  The mean-square displacement $\langle x^2_N\rangle $ is simply the mean of Eq. (\ref{eq:Diffusion:X-sum}) squared:
\begin{equation}\label{eq:Diffusion:MSD_sum}
\langle x_N^2\rangle  = \langle (\Delta x_1 + \Delta x_2 + ... + \Delta x_N)^2\rangle,
\end{equation}
(in this work $\langle\ \rangle$ denotes the mean). As long as the diffusion is nonanomalous (free of advection) the average random walker position is $\langle x_N \rangle=0$. Thus, the mean-square displacement of a single random walker is the same as the variance for the distribution of random walker positions. Using an elementary arithmetic rule for random variables, Eq. (\ref{eq:Diffusion:MSD_sum}) may be rewritten:
\begin{equation}\label{eq:Diffusion:CovarianceMatrix}
\langle x_N^2\rangle  = \sum_{j=1}^{N}\left( \sum_{k=1}^{N} \langle \Delta x_j \Delta x_k  \rangle \right).
\end{equation}
As $\langle \Delta x_i\rangle =0$, the expectation value $ \langle \Delta x_j \Delta x_k  \rangle $ is the covariance of the two random variables, $\Delta x_j$ and $\Delta x_k$, and Eq. (\ref{eq:Diffusion:CovarianceMatrix}) is simply the sum of the corresponding covariance matrix. Clearly, the absolute values of the indices $j$ and $k$ are irrelevant, only their absolute difference matters. This significantly simplifies the calculation of the covariance matrix, which has the elements:
\begin{equation}\label{eq:covarmatrix}
\Sigma_{jk}=\mathrm{cov}(\Delta x_j, \Delta x_k)= \langle \Delta x_i \Delta x_{i+m}\rangle,
\end{equation}
where $m=|k-j|$, (\ie, the covariance matrix is a symmetric Toeplitz matrix).\\

Considering this random walk projected onto an axis as a 1D diffusion problem it is easy to show that:
\begin{equation} \label{eq:Dxxt}
\langle x^2(t)\rangle  = 2D_{xx}t.
\end{equation}
After many steps (which is required for diffusion theory to be valid) it may safely be assumed that $t=N\langle \Delta t\rangle $, where $\langle \Delta t\rangle $ is the mean of the step time distribution. Hence, Eq. (\ref{eq:Dxxt}) can be converted from continuous time ($t$)  to discrete time ($N$ steps):
\begin{equation}\label{eq:Diffusion:DiffVariance}
D_{xx} = \frac{\langle x^2_N\rangle }{2N\langle \Delta t\rangle }.
\end{equation}\
In conclusion, the diffusion tensor element can be calculated by summing the elements of the covariance matrix, Eq. (\ref{eq:covarmatrix}). This method provides an intuitive understanding of the connection between the random walk and the diffusion tensor and can, for example, analytically re-create the result for isotropic diffusion, Eq. (\ref{eq:Diso}); see Ref. \cite{Alerstam2011_PhDThesis} for details.\\

In cases where $D_{xx}$ lacks an obvious or simple closed form expression, it may be calculated numerically. This can be performed with a computationally inexpensive random walk simulation, where a single random walker is traced for a large number of steps in an unbounded absorptionless three-dimensional medium with the desired differential scattering cross section. For each step, $i$, the projections of the step onto each of the principal axes radiance ellipsoid (in this case, the axis of the cartesian coordinate system$\Delta x_i$, $\Delta y_i$, and $\Delta z_i$) are stored, along with the time it took to complete the step, $\Delta t_i$. After the simulation, the elements of the covariance matrix may be calculated using
\begin{equation}
 \langle \Delta x_i \Delta x_{i+m}\rangle = \frac{1}{N-m}\sum\limits_{i=1}^{N-m} \Delta x_i \Delta x_{i+m}.
\end{equation}
Due to the symmetric nature of the covariance matrix, its sum [Eq. \ref{eq:Diffusion:CovarianceMatrix}] may be calculated using
\begin{equation} \label{eq:covarmatrix_sum}
\langle x_N^2\rangle \stackrel{N\gg1}{\approx} N\left( \langle \Delta x_i \Delta x_{i}\rangle + 2\sum\limits_{m=1}^{N} \langle \Delta x_i \Delta x_{i+m}\rangle\right).
\end{equation}
Here it was assumed that $N$ is large (it is worth repeating that this assumption is a prerequisite for diffusion theory to be applicable in the first place). Note that ordinarily $ \langle \Delta x_i \Delta x_{i+m}\rangle$ approaches zero as $m$ becomes large because correlation between the steps is gradually lost due to the randomness in the scattering. This reduces the computation required of Eq. (\ref{eq:covarmatrix_sum}) as convergence is achieved when $ \langle \Delta x_i \Delta x_{i+m}\rangle$ reaches a sufficiently small value.

Finally, the average step time, $\langle\Delta t_i\rangle$, is calculated and inserted into Eq. (\ref{eq:Diffusion:DiffVariance}) along with Eq. (\ref{eq:covarmatrix_sum}) to yield the diffusion tensor element:
\begin{equation} \label{eq:D_RWT}
D_{xx}^{\mathrm{RWT}} = \frac{\frac{1}{2}\langle \Delta x_i \Delta x_{i}\rangle + \sum\limits_{m=1}^{N} \langle \Delta x_i \Delta x_{i+m}\rangle}{\langle \Delta t_i \rangle}.
\end{equation}

\subsection{Deriving the extrapolation length for anisotropic radiance}\label{sec:ze}
In addition to an appropriate diffusion tensor, diffusion modeling of bounded media also requires an appropriate boundary condition. Equation (\ref{eq:I_tr}) utilizes the so-called extrapolated boundary condition \cite{Keijzer1988_ApplOpt,Zhu1991_PRA,Contini1997_ApplOpt}, where the fluence is extrapolated to zero at a virtual boundary located a small distance outside the actual boundary. The distance to this boundary, called the extrapolation length $z_e$, is usually derived assuming an isotropic radiance. However, as discussed in Sec. \ref{sec:anisoRadiance}, for macroscopically anisotropic systems, an isotropic radiance can no longer be assumed. The appendix outlines the modification of the extrapolated boundary condition to accommodate an anisotropic radiance. In order to be applicable, the modified boundary condition requires knowledge of the stationary angular distribution of the radiance, $P(\mathbf{\hat{s}})$. As discussed in Sec. \ref{subsec:aniso_radiance}, and Fig. \ref{fig:Radiance}, the stationary distribution can be calculated using the Monte Carlo method to solve the Markov Chain problem. Alternatively, one may recognize that the directions $\mathbf{s}_i=[\Delta x_i,\Delta y_i,\Delta z_i]$ obtained from the Monte Carlo simulation described in the calculation of the diffusion tensor (Sec. \ref{sec:diffusion_tensor}) also represent the stationary angular distribution as long as the number of steps, $N$, is large. Here, the distribution function $P(\mathbf{\hat{s}})$ is estimated by the distribution of directions: $\mathbf{\hat{s}}_i$ weighted by the length of each step, $s_i = \sqrt{(\Delta x_i)^2+(\Delta y_i)^2+(\Delta z_i)^2}$. The direction vector $\mathbf{\hat{s}}_i$ is a unit vector, \ie, $\mathbf{\hat{s}}_i = [\Delta x_i,\Delta y_i,\Delta x_i]/s_i = [dx_i\ dy_i\ dz_i]$. Further, assuming a boundary with a normal $\mathbf{\hat{n}}$ aligned with the $z$ axis, gives $\cos\theta_i = |dz_i|$. The coefficients $A$, $B$, $X$, and $Y$ [Eqs. (\ref{eq:B}), (\ref{eq:C}), (\ref{eq:X}), and (\ref{eq:Y})], can hence be estimated numerically:
\begin{align}
B = \frac{1}{N\langle s_i\rangle}&\sum_{i=1}^N \cos\theta_iv(\mathbf{\hat{s}}_i)s_i,\\
C = \frac{1}{N\langle s_i\rangle}&\sum_{i=1}^N \cos^2\theta_i\ell^\ast (\mathbf{\hat{s}}_i)v(\mathbf{\hat{s}}_i)s_i,\\
X = \frac{1}{N\langle s_i\rangle}&\sum_{i=1}^N \cos\theta_iv(\mathbf{\hat{s}}_i)s_iR_F(\theta_i),\\
Y = \frac{1}{N\langle s_i\rangle}&\sum_{i=1}^N \cos^2\theta_i\ell^\ast (\mathbf{\hat{s}}_i)v(\mathbf{\hat{s}}_i)s_iR_F(\theta_i),
\end{align}
where $\ell^\ast$ is calculated using Eq. (\ref{eq:ellstar}). This gives an expression for the extrapolation length:
\begin{align}\label{eq:ze_RWT}
z_{e}^{\mathrm{RWT}} &= \frac{\sum_{i=1}^N \cos^2\theta_i\ell^\ast (\mathbf{\hat{s}}_i)v(\mathbf{\hat{s}}_i)s_i(1-R_F(\theta_i))}{\sum_{i=1}^N \cos\theta_iv(\mathbf{\hat{s}}_i)s_i(1-R_F(\theta_i))}.
\end{align}

\subsection{Numerical estimation of $\mathbf{D}$ and $z_e$}\label{subsec:Numest}
A separate CPU-based code was implemented to perform the random walk of a single walker in an unbounded medium. A direction-dependent scattering coefficient, $\mu_s(\mathbf{\hat{s}})$, and anisotropy factor $g(\mathbf{\hat{s}})$ was implemented, as described in Sec. \ref{sec:MC}. The data from the random walk was directed to a subroutine, which performed the calculation of the elements of the diffusion, $\mathbf{D}^\mathrm{RWT}$ [Eq. (\ref{eq:D_RWT})] and the extrapolation length, $z_{e}^{\mathrm{RWT}}$ [Eq. (\ref{eq:ze_RWT})]. The number of steps for the random walker was set to $N=5\times10^6$. For all the presented results, this procedure was repeated 50 times in order to give a good estimate of the mean and standard deviation of the calculated quantities, and to reduce the effect of stochastic noise.\\

In a first test, the numerical methods and implementations for the calculation of the diffusion tensor and extrapolation length were tested using isotropic systems, where the results are well known. For example, using the Henyey-Greenstein phase scattering function with $g=0.8$, $\mu_s = 100$ cm$^{-1}$, and $n=1.4$, the numerical algorithm produced $D^\mathrm{RWT}=35.7\pm0.2$ m$^2$/ms and $z_{e}^{\mathrm{RWT}}=0.984\pm0.003$ mm (mean $\pm$ sample standard deviation). These numbers are in good agreement with the results from analytical theory: $D^\textrm{iso} = 35.7$ m$^2$/ms and $z_{e}^{\mathrm{iso}} = 0.983$ mm, using Eq. (\ref{eq:Diso}), and the method of Contini \etal\ \cite{Contini1997_ApplOpt}, respectively.

\section{A discussion on the relation between $\mathbf{D}$, $\boldsymbol{\mu}_s$ and $\mathbf{g}$}\label{sec:Dmusg}
\begin{figure}
\begin{center}
  \includegraphics{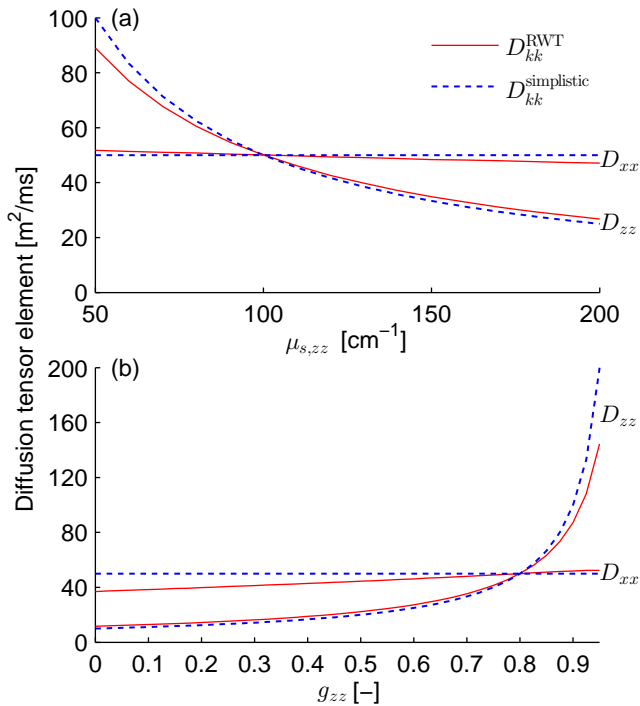}
 \caption{(Color online) Diffusion tensor elements calculated using random walk theory (RWT) [Eq. (\ref{eq:D_RWT})] and the simplistic method [Eq. (\ref{eq:Dsimplistic})] as a function of a single varying microscopic scattering property. In (a) $\mu_{s,zz}$ is varied while $g_{xx}=g_{yy}=g_{zz}=0.8$, and $\mu_{s,xx}=\mu_{s,yy} = 100$ cm$^{-1}$ are kept constant. In (b) $g_{zz}$ is varied, and $g_{xx}=g_{yy}=0.8$, and $\mu_{s,xx}=\mu_{s,yy}=\mu_{s,zz} = 100$ cm$^{-1}$ are kept constant. In both (a) and (b), the refractive index is $n=1.0$ and $\mu_a=0$. Due to symmetry, $D_{yy}=D_{xx}$. It is clear that the simplistic method does not provide an accurate estimate of the diffusion tensor, unless the diffusion is isotropic.}\label{fig:D_vs_musz_gz}
\end{center}
\end{figure}

Diffusion tensor elements for a wide range of microscopic scattering properties were calculated using the simplistic method [Eq. (\ref{eq:Dsimplistic})] and compared to the accurate random walk theory (RWT) [Eq. (\ref{eq:D_RWT})]. The results of two test cases are illustrated in Fig. \ref{fig:D_vs_musz_gz}. The baseline optical properties were: $g_{xx}=g_{yy}=g_{zz}=0.8$, $n=1.0$, $\mu_{s,xx}=\mu_{s,yy}=\mu_{s,zz} = 100$ cm$^{-1}$, and $\mu_a=0$. In (a) $\mu_{s,zz}$ is varied from 50 to 200 cm$^{-1}$ while in (b) $g_{zz}$ is varied between 0.0 and 0.95.

When the diffusion is isotropic: $\mu_{s,zz}=100$ cm$^{-1}$ in (a) and $g_{zz} = 0.8$ in (b), both methods reach the same results. However, as discussed in Sec. \ref{sec:diffusion_tensor}, the relationship between the diffusion tensor and the microscopic scattering properties is not trivial when the diffusion is anisotropic. As expected the simplistic method clearly fails to reproduce the results of the accurate RWT. While this result may seem trivial, it has several interesting and important consequences. First, it is noted that changing, for example, the scattering coefficient in one direction changes all elements in $\mathbf{D}$. This explains an apparent inconsistency in anisotropic diffusion theory: It is intuitively obvious that the transmission through a slab is dependent on the scattering properties in all directions, yet Eq. (\ref{eq:I_tr}) implies that the transmission straight through a slab, $I_{tr}(0,0,t)$, is independent of $D_{xx}$ and $D_{yy}$. Figure \ref{fig:D_vs_musz_gz} clearly illustrates that $D_{zz}$ encompasses the scattering properties in all directions, not only the scattering properties in the $z$ direction. Thus, changing the microscopic scattering properties in the direction perpendicular to the $z$ axis will be represented by a change in $D_{zz}$. Furthermore, there is no simple (inverse) proportionality between the elements in $\mathbf{D}$ and the microscopic scattering properties. For example, the ratio $D_{xx}/D_{zz}$ in Fig. \ref{fig:D_vs_musz_gz} (a) reveals a dependence on the single-scattering anisotropy factor, $g=g_{xx}=g_{yy}=g_{zz}$ (data not shown). Consequently, little or no reliable information about the microscopic scattering properties can be deduced from diffusion tensor element ratios without \textsl{a priori} information regarding the microstructure.

\section{Agreement between Monte Carlo simulations and anisotropic diffusion theory}\label{sec:diffvsMC}
\subsection{Steady state}
Spatially resolved transmission through a slab was chosen as a test case for steady-state solutions of the anisotropic diffusion equation:
\begin{equation}
I_\mathrm{tr}(x,y) = \int\limits_{t=0}^\infty I_\mathrm{tr}(x,y,t)dt.
\end{equation}
Studying the transmission through an optically thick slab ensures that the diffusion approximation is valid (which is not the case for reflectance measurements close to the source). Any observed discrepancy in the comparison to the results of Monte Carlo simulations is thus due to invalidity of the anisotropic diffusion theory, not due to a breakdown of diffusion theory in general. Monte Carlo simulations were performed for non-absorbing slabs, with varying thicknesses, refractive indices, and for a wide variety of anisotropic as well as isotropic (for validation) microstructures. For comparison, diffusion tensors were calculated with the presented random walk theory $\mathbf{D}^\mathrm{RWT}$ [Eq. (\ref{eq:D_RWT})] as well as using the simplistic method, $\mathbf{D}^\mathrm{simplistic}$ [Eq. (\ref{eq:Dsimplistic})]. Extrapolation lengths were calculated using the theory modified for anisotropic radiance, $z_{e}^{\mathrm{RWT}}$ [Eq. (\ref{eq:ze_RWT})], as well as the conventional method assuming isotropic radiance, $z_{e}^{\mathrm{iso}}$ \cite{Contini1997_ApplOpt}. In all cases, $z_0$ was calculated using Eq. (\ref{eq:Dzzell}).
\begin{figure*}
\begin{center}
  \includegraphics{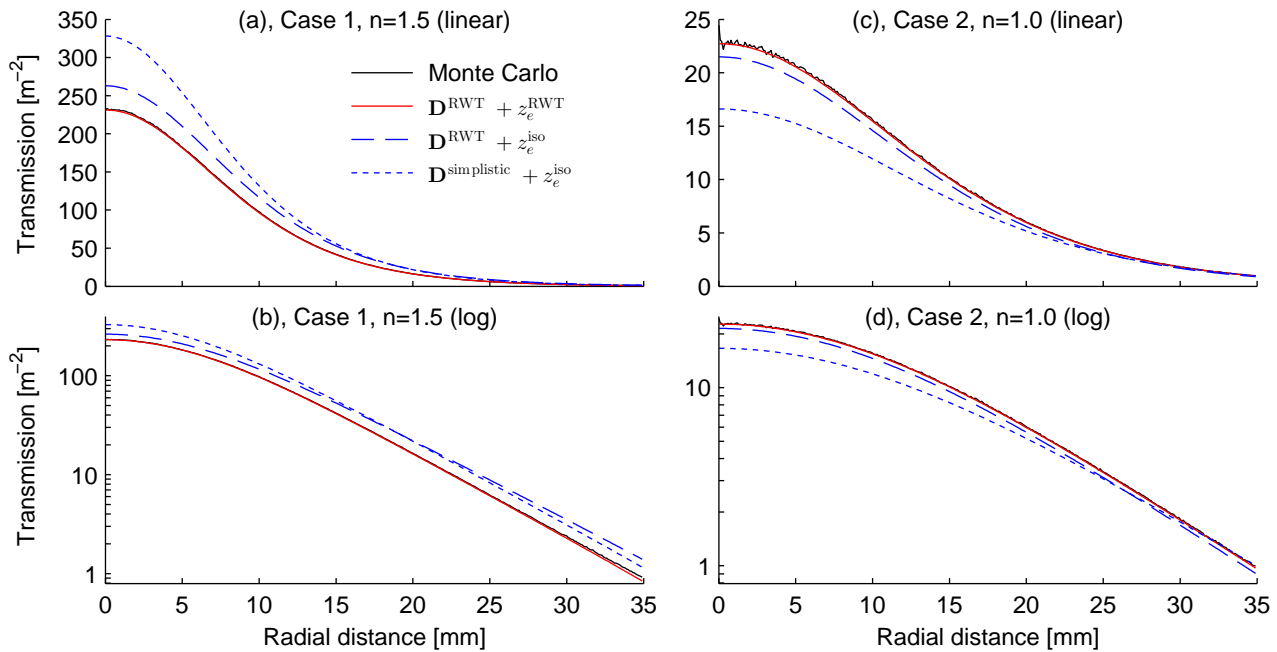}
 \caption{(Color online) Steady-state transmission through 2-cm-thick, nonabsorbing slabs, as modeled by Monte Carlo and anisotropic diffusion theory. (a) and (b) show Case 1 (in linear and logarithmic, scale respectively), where the anisotropy is due to a direction-dependent scattering coefficient, $\mu_{s,xx}=\mu_{s,yy} = 100$ cm$^{-1}$, and $\mu_{s,zz} = 50$ cm$^{-1}$, with $g_{xx}=g_{yy}=g_{zz}=0.8$, and $n=1.5$. The panels to the right (c) and (d) show Case 2 (in linear and logarithmic, scale respectively), where the single-scattering anisotropy factor is direction-dependent with $g_{xx}=g_{yy}=0.8$, and $g_{zz}=0.6$, while $n=1.0$, and $\mu_{s,xx}=\mu_{s,yy}=\mu_{s,zz} = 100$ cm$^{-1}$. Using RWT, anisotropic diffusion reproduce the Monte Carlo results within the accuracy of diffusion theory. However, using the simplistic method for deriving the diffusion tensor, and/or using the boundary condition for isotropic radiance, results in large deviations compared to Monte Carlo.}\label{fig:SS_radial}
\end{center}
\end{figure*}

Two test cases, Case 1 and 2, with refractive indices of 1.5 and 1.0, respectively, were selected as representative results and are shown in Fig. \ref{fig:SS_radial}. These two cases provide solutions that are rotationally symmetric around the $z$-axis. Hence, Fig. \ref{fig:SS_radial} shows the transmission as a function of the radial distance from the $z$-axis, a configuration that greatly increases the signal-to-noise in the Monte Carlo simulation results. The inadequacy of the simplistic method is obvious, while anisotropic diffusion using the RWT-derived diffusion tensor and the boundary condition modified for anisotropic radiance accurately models the transmission. In all cases, the deviation in absolute transmission between Monte Carlo and diffusion (combined with RWT derived extrapolation length and diffusion tensor) was generally $<1\%$ with the worst observed deviation of $\approx3\%$. This variation is well within the limits the accuracy of the extrapolated boundary condition method (see, e.g., Ref. \cite{Contini1997_ApplOpt}). A small dependency on the radial distance was observed in the deviation in some cases [for example, Fig. \ref{fig:SS_radial}(b)], which is possibly attributed to a deviation from the stationary radiance distribution far away from $r=0$ and close to the boundary.

Also illustrated in Fig. \ref{fig:SS_radial} are the diffusion results where $\mathbf{D}^\mathrm{RWT}$ is combined with the conventional extrapolated boundary condition assuming isotropic radiance, $z_{e}^{\mathrm{iso}}$. This isolates the effect of using the improper boundary condition, illustrating the importance of the modification to account for the anisotropic radiance, as described earlier.  A notable result, illustrated in Figs. \ref{fig:SS_radial}(c)-\ref{fig:SS_radial}(d), is that the conventional, isotropic, extrapolated boundary condition is inaccurate, even when in the absence of a refractive index boundary mismatch ($n=n_0=1.0$). This is predicted by Eq. (\ref{eq:ze1}), where $X$ and $Y$ are zero due to their dependance on the Fresnel reflection coefficient, $R_F(\theta)$, while $B$ and $C$ are still dependent on the angular distribution of the radiance.

It is also worth noting that using $z_{e}^{\mathrm{iso}}$ leads to an over- and underestimation compared to the Monte Carlo transmission results respectively for the two test cases. This illustrates how the modified boundary condition is dependent on the shape of the anisotropic radiance distribution, $P(\mathbf{\hat{s}})$. For example, in Case 1, the random walkers are more likely to travel in the direction perpendicular to the boundaries, causing the effective reflection coefficient to decrease relative to the isotropic radiance situation. Compared to assuming isotropic radiance, this cause more losses through the boundary close to the source which in turn decreases overall transmission, as observed in Figs. \ref{fig:SS_radial}(a)-\ref{fig:SS_radial}(b).

\subsection{Time domain}
\begin{figure*}
\begin{center}
  \includegraphics{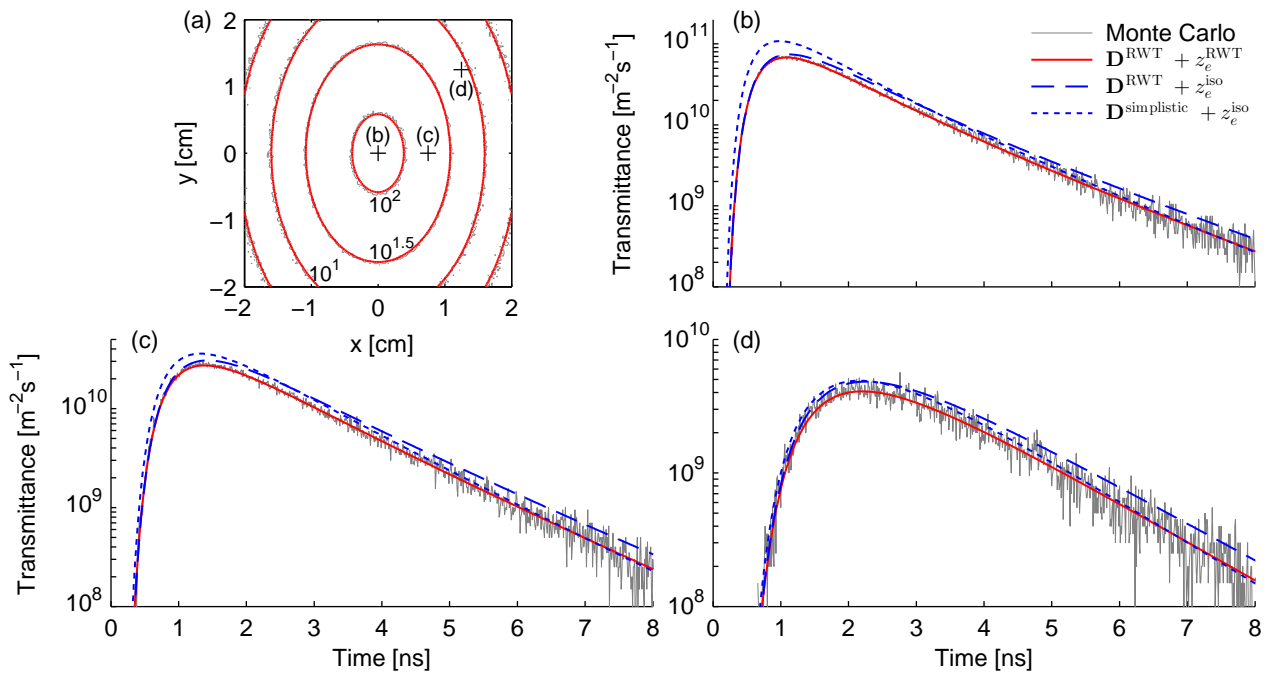}
  \caption{(Color online) (a) Isointensity curves for the steady-state transmission (units: [m$^{-2}$]) through a 2-cm-thick slab with optical properties: $\mu_{s,xx}=\mu_{s,yy}=\mu_{s,zz} = 100$ cm$^{-1}$, $g_{xx}=g_{yy}=0.8$, $g_{zz}=0.4$, $\mu_a=0$, and $n=1.5$.  Panels (b)-(d) show the time-resolved transmittance through the slab at the positions marked in (a). Each plot shows the results of a Monte Carlo simulation (random walkers collected from a $2\times 2$ mm$^2$ area centered around the collection coordinates), as well as results from anisotropic diffusion theory. Diffusion theory with the RWT-derived diffusion tensor and extrapolation length show excellent agreement with Monte Carlo results, while using the simplistic diffusion tensor and/or the isotropic boundary condition results in large deviations.}\label{fig:TOF_transmission}
\end{center}
\end{figure*}

The temporally resolved transmittance through slabs with a wide variety of thicknesses and optical properties (refractive index, and $\boldsymbol{\mu}_s$ and $\mathbf{g}$ tensors) were studied at various $x$ - $y$ positions by comparing the results of Monte Carlo simulations and anisotropic diffusion theory. Overall, excellent agreement was found when using the RWT method for deriving the diffusion tensor and the extrapolation length, while the simplistic diffusion tensor element method and the isotropic radiance boundary condition both introduced significant errors. To illustrate these results, Figs. \ref{fig:TOF_transmission}(b)-\ref{fig:TOF_transmission}(d) show the temporally resolved transmittance through a 2-cm-thick slab with nonrotationally symmetric anisotropy, at various positions. Figure \ref{fig:TOF_transmission}(a) illustrates the steady-state isointensity curves for the same slab, as well as the positions of collection for Figs. \ref{fig:TOF_transmission}(b)-\ref{fig:TOF_transmission}(d).

As expected, anisotropic diffusion theory (using RWT for diffusion tensor and extrapolation length calculation) is in good agreement with Monte Carlo simulations, while the simplistic diffusion tensor calculation and the isotropic boundary condition both introduce significant deviations. As in the steady-state comparisons between RWT diffusion and Monte Carlo results, small (on the order of one percent), offset-like deviations were observed in some comparisons. This can most likely be attributed to the limited accuracy of the approximative boundary condition. 

Studying Figs. \ref{fig:TOF_transmission}(b)-\ref{fig:TOF_transmission}(c) it appears like diffusion theory using $\mathbf{D}^\mathrm{simplistic}$, in combination with $z_{e}^{\textrm{iso}}$, produces results that converge toward the Monte Carlo solution at later times (approximately $t>$ 6 ns). This behavior was not consistently observed for other simulations with other optical properties. It is thus important to note that this purely accidental behavior cannot be used, for example, for extraction of absorption properties. In general, in order to accurately extract the absorption coefficient through the fitting of time-resolved data with anisotropic diffusion theory, a correct boundary condition is required, even if the diffusion tensor is configured to freely vary during the fitting procedure. This is illustrated in Figs. \ref{fig:TOF_transmission} (b)-\ref{fig:TOF_transmission}(d). Here, the combination of the RWT-derived diffusion tensor with an inappropriate extrapolation length leads to large deviations from the Monte Carlo solution.

\section{Discussion and conclusions}\label{sec:discussion}
The presented theory provides an accurate and intuitive method to calculate the diffusion tensor for systems of independent scatterers with arbitrary microscopic scattering properties. Here, the theory is tested against Monte Carlo simulations using an unphysical test case where the microscopic scattering properties are separated into two independent tensors ($\boldsymbol{\mu}_s$ and $\mathbf{g}$). However, the theory is applicable to arbitrary systems, such as the physically more realistic case of statistically aligned anisotropic  scattering cylinders. The presented method utilizes the strength of the Monte Carlo method where scatterers with an arbitrary differential scattering cross section, statistical alignment, and/or a mix of several types of scatterers, is easily implemented. Furthermore, like the regular Monte Carlo method, the presented numerical method can be extended to account for polarization effects such as polarization-dependent scattering. Contrary to the computationally costly method of using Monte Carlo to solve the transport problem for the entire system, the calculation of the diffusion tensor only requires a single random walker to be tracked for a few million steps, an operation of negligible computational cost for a modern computer. The computation required for summing the elements of the covariance matrix is more expensive, but is also a task that is easily parallelizable. It should also be noted that the theory can be used to derived analytical relationships for specific systems (see Ref. \cite{Alerstam2011_PhDThesis} for systems with randomly oriented anisotropic scatterers, \ie, isotropic diffusion).

The results presented in this work clearly show that, contrary to previous studies \cite{Kienle2007_PRL,Kienle2013_PhysMedBiol}, anisotropic diffusion theory indeed is valid (within the limitations of diffusion theory). The fallacy that lead to the conclusion that anisotropic diffusion is invalid was the use of an invalid boundary condition and, as also pointed out by Johnson \etal, an invalid relationship between the microscopic scattering properties and the diffusion tensor \cite{Johnson2009_JBO}. As shown in this work, the relationship between the microscopic scattering properties and the diffusion tensor is complicated and dependent on the specific microscopic scattering properties of the system under consideration. The presented theory intuitively connects the micro- and macroscopic properties by studying the statistics (\ie, the covariance matrix) of steps taken in a random walk. The off-diagonal elements of the covariance matrix represent long-range step correlations, which in turn arise from a direction-dependent single-scattering phase function, and/or an anisotropic single-scattering phase function (such as the direction independent Henyey-Greenstein function). Clearly, this stands in conflict with the method to calculate the diffusion tensor presented by Johnson \etal\ \cite{Johnson2009_JBO}, which doesn't take the step correlations or the anisotropic radiance into account. 

This work also presents a boundary condition modification to account for an anisotropic radiance. When the transport anisotropy is fully or partly due to a direction-dependent phase scattering function, the stationary radiance direction distribution is not trivial to derive, as it takes multiple steps for the radiance to converge. Using the presented RWT method, the steps taken by the single walker can be used as a representative sample of the stationary radiance distribution, from which the extrapolation length can be numerically calculated. It should be noted that the extrapolated boundary condition is approximative and does lead to small deviations between Monte Carlo simulations and diffusion theory, even in the case of isotropic diffusion in a regime where diffusion theory is a very good approximation. The small mismatches observed in this work are comparable to the boundary-condition-induced mismatches observed when comparing Monte Carlo solutions with isotropic diffusion.

The requirement of a boundary condition modified for anisotropic radiance poses a challenge to the applicability of anisotropic diffusion theory in practice. The numerical calculation of the diffusion tensor and the extrapolation length can be made fast enough to be useful in e.g., iterative curve fitting of diffusion models with experimental, or simulation data. Thus, it can be useful when the microscopic nature of the scatterers is known and the appropriate differential scattering cross section can be applied in the forward model. 
In the absence of this \textsl{a priori} information, the fitting of solutions to the anisotropic diffusion equation to experimental, or simulation, data, is less straightforward. Accurate diffusion tensors or absorption coefficients can only be extracted as long as an appropriate extrapolation length is used. However, as shown in this work, the extrapolation length is dependent on the radiance, which is calculated from the microscopic optical properties. Still, owing to the link between the diffusion tensor, the radiance, and the extrapolation length, it is not unreasonable to imagine that an approximate relationship between the extrapolation length and the diffusion tensor can be found (as it has for isotropic diffusion; see e.g., Ref. \cite{Contini1997_ApplOpt}). Such a relationship would be helpful in studies of systems with anisotropic transport using diffusion theory, and is a recommended direction of future work.

\section*{Acknowledgments}
Part of this work was supported and generously funded by Stefan Andersson-Engels using grants from the Swedish Research Council. Tomas Svensson is gratefully acknowl- edged for interesting discussions, constant encouragement, and invaluable help improving the manuscript. The author is also grateful to Kevin Vynck, Matteo Burresi, and Corey Cochrane for reading and helping to improve the manuscript. 

Lorenzo Pattelli is gratefully acknowledged for bringing a mistake in Eq. (\ref{eq:I_tr}) to my attention. An erratum has been submitted and this version has been updated to correct the mistake in the printed version of the paper.

\appendix
\section{Boundary condition}\label{appendix:boundarycondition}
The treatment of boundaries with refractive index mismatch for turbid media have been extensively investigated over the years by several authors (e.g., Refs. \cite{Keijzer1988_ApplOpt,Zhu1991_PRA,Contini1997_ApplOpt}). The extrapolated boundary condition has emerged as a popular method to handle boundaries. However, in the derivation of the extrapolation length, $z_e$, it is commonly assumed that the radiance is near isotropic, a condition that fails for systems with anisotropic transport properties. Here, the extrapolated boundary condition, as derived by Zhu \etal\ \cite{Zhu1991_PRA}, is modified to accommodate an anisotropic radiance.\\

The flux out through the boundary, $J_-$, is:
\begin{align}
J_-= &\int\limits_0^{\pi/2} d\theta \int\limits_0^{2\pi}d\phi \int\limits_0^{\infty} dr \frac{L(\mathbf{r},\mathbf{\hat{s}}) \sin(\theta) \cos(\theta)}{\ell^{\ast}(\mathbf{\hat{s}})} \nonumber \\
& \times \exp\left(-\frac{r}{\ell^{\ast}(\mathbf{\hat{s}})}\right), \label{eq:J-}
\end{align}
where the direction-dependent transport mean free path, $\ell^{\ast}(\mathbf{\hat{s}})$, is given by:
\begin{equation}\label{eq:ellstar}
\ell^{\ast}(\mathbf{\hat{s}}) = \frac{3D(\mathbf{\hat{s}})}{v(\mathbf{\hat{s}})}=\frac{3\mathbf{\hat{s}}\mathbf{D}\mathbf{\hat{s}}^{\mathrm{T}}}{v(\mathbf{\hat{s}})}.
\end{equation}

Inserting Eq. (\ref{eq:radiance}) and linearizing the energy density, $W(\mathbf{r})$, around $z=0$ using a first-order Taylor expansion gives:
\begin{equation}\label{eq:Taylor}
W(\mathbf{r}) = W_0 + x\left.\frac{\partial W}{\partial x}\right|_{z=0} + y\left.\frac{\partial W}{\partial y}\right|_{z=0} + z\left.\frac{\partial W}{\partial z}\right|_{z=0}.
\end{equation}
Transformation from cartesian to spherical coordinates is done by:
\begin{eqnarray}
x &=& r \sin \theta \cos \phi \nonumber \\
y &=& r \sin \theta \sin \phi \label{eq:cart2sph}\\
z &=& r \cos \theta.  \nonumber
\end{eqnarray}
Equations (\ref{eq:Taylor}) and (\ref{eq:cart2sph}) can now be inserted into Eq. (\ref{eq:J-}). Just like in the isotropic case, the terms containing $x$ and $y$ vanishes due to the integration of $\phi$ over $2\pi$. This assumes that both $v(\mathbf{\hat{s}})$ and $P(\mathbf{\hat{s}})$ are well-behaved, smooth, and symmetric (or antisymmetric) functions. For $P$, these constraints are the same as for diffusion theory to be valid in the first place, i.e., when plotted in spherical coordinates $P$ must be represented by a spheroid (cf. Fig. \ref{fig:Radiance}).

The resulting fluxes out ($J_-$) and in ($J_+$) through the boundary (at $z=0$) are:
\begin{align}
J_\pm = BW_0 \mp C\left.\frac{\partial W}{\partial z}\right|_{z=0},
\end{align}
where
\begin{align}
B&=& \int\limits_0^{\pi/2} d\theta \int\limits_0^{2\pi}d\phi &P(\mathbf{\hat{s}})v(\mathbf{\hat{s}}) \sin(\theta) \cos(\theta),\label{eq:B}\\
C&=& \int\limits_0^{\pi/2} d\theta \int\limits_0^{2\pi}d\phi &P(\mathbf{\hat{s}})v(\mathbf{\hat{s}}) \sin(\theta) \cos^2(\theta)\ell^{\ast}(\mathbf{\hat{s}}).\label{eq:C}
\end{align}
$J_-$ and $J_+$ are related through
\begin{equation}
J_+ = R_\mathrm{eff}J_-,
\end{equation}
where $R_\mathrm{eff}$ is the effective reflection coefficient. This gives:
\begin{align}
W_0 - \frac{C(1+R_\mathrm{eff})}{B(1-R_\mathrm{eff})}\left. \frac{\partial W}{\partial z}\right|_{z=0} = 0.
\end{align}
Linearly extrapolating the energy density to zero gives the extrapolation length \cite{Zhu1991_PRA}
\begin{equation}\label{eq:ze1}
z_e =  \frac{C(1+R_\mathrm{eff})}{B(1-R_\mathrm{eff})} = \frac{C+Y}{B-X},
\end{equation}
where
\begin{align}
X &=& \int\limits_0^{\pi/2} d\theta \int\limits_0^{2\pi}d\phi &P(\mathbf{\hat{s}})v(\mathbf{\hat{s}}) \sin(\theta) \cos(\theta) R_\mathrm{F}(\theta),\label{eq:X}\\
Y &=& \int\limits_0^{\pi/2} d\theta \int\limits_0^{2\pi}d\phi &P(\mathbf{\hat{s}})v(\mathbf{\hat{s}}) \sin(\theta) \cos^2(\theta)\ell^{\ast}(\mathbf{\hat{s}})R_\mathrm{F}(\theta).\label{eq:Y}
\end{align}
Here, $R_\mathrm{F}(\theta)$, is the reflection coefficient for an incident angle $\theta$, calculated according to Fresnel's law, averaged over polarization. This gives $z_e$ for any given $P(\mathbf{\hat{s}})$, $v(\mathbf{\hat{s}})$, and $\mathbf{D}$.


\begin{thebibliography}{10}

\bibitem{Kao1996_PRL}
M.~H. Kao, K.~A. Jester, A.~G. Yodh, and P.~J. Collings.
\newblock Observation of light diffusion and correlation transport in nematic
  liquid crystals.
\newblock {\em Phys. Rev. Lett.} \textbf{77}, 2233--2236 (1996).

\bibitem{Stark1996_PRL}
H.~Stark and T.~C. Lubensky.
\newblock Multiple light scattering in nematic liquid crystals.
\newblock {\em Phys. Rev. Lett.} \textbf{77}, 2229--2232 (1996).

\bibitem{VanTiggelen1996_PRL}
B.~A. van Tiggelen, R.~Maynard, and A.~Heiderich.
\newblock Anisotropic light diffusion in oriented nematic liquid crystals.
\newblock {\em Phys. Rev. Lett.}, \textbf{77}, 639--642, (1996).

\bibitem{Johnson2002_PRL}
P.~M. Johnson, B.~P.~J. Bret, J.~G\'{o}mez~Rivas, J.~J. Kelly, and
  A.~Lagendijk.
\newblock Anisotropic diffusion of light in a strongly scattering material.
\newblock {\em Phys. Rev. Lett.} \textbf{89} 1--4, (2002).

\bibitem{Johnson2008_OptExpress}
P.~M. Johnson, S.~Faez, and A.~Lagendijk.
\newblock Full characterization of anisotropic diffuse light.
\newblock {\em Opt. Express} \textbf{16} 7435--7446, (2008).

\bibitem{Johnson2009_JBO}
P.~M. Johnson and A.~Lagendijk.
\newblock Optical anisotropic diffusion: new model systems and theoretical
  modeling.
\newblock {\em J. Biomed. Opt} \textbf{14} 054036, (2009).

\bibitem{Tsuchikawa2002_ApplSpectrosc}
S.~Tsuchikawa and S.~Tsutsumi.
\newblock Application of time-of-flight near-infrared spectroscopy to wood with
  anisotropic cellular structure.
\newblock {\em Appl. Spectrosc.} \textbf{56} 869--876, (2002).

\bibitem{Kienle2008_OptExpr}
A.~Kienle, C.~D'Andrea, F.~Foschum, P.~Taroni, and A.~Pifferi.
\newblock Light propagation in dry and wet softwood.
\newblock {\em Opt. Express} \textbf{16} 9895--9906, (2008).

\bibitem{Nickell2000_PhysMedBiol}
S.~Nickell, M.~Hermann, M.~Essenpreis, Farrell~T. J., U.~Kr\"{a}mer, and M.~S.
  Patterson.
\newblock Anisotropy of light propagation in human skin.
\newblock {\em Phys. Med. Biol.} \textbf{45} 2873--2886, (2000).

\bibitem{Sviridov2005_JBO}
A.~Sviridov, V.~Chernomordik, M.~Hassan, A.~Russo, A.~Eidsath, P.~Smith, and
  A.~Gandjbakhche.
\newblock Intensity profiles of linearly polarized light backscattered from
  skin and tissue-like phantoms.
\newblock {\em J. Biomed. Opt.} \textbf{10} 014012, (2005).

\bibitem{Kienle2006_PRL}
A.~Kienle and R.~Hibst.
\newblock Light guiding in biological tissue due to scattering.
\newblock {\em Phys. Rev. Lett.} \textbf{97} 018104, (2006).

\bibitem{Alerstam2012_PRE}
E.~Alerstam and T.~Svensson.
\newblock Observation of anisotropic diffusion of light in compacted granular
  porous materials.
\newblock {\em Phys. Rev. E} \textbf{85} 040301, (2012).

\bibitem{vanTiggelen2000_RevModPhys}
B.~van Tiggelen and H.~Stark.
\newblock Nematic liquid crystals as a new challenge for radiative transfer.
\newblock {\em Rev. Mod. Phys.} \textbf{72} 1017--1039, (2000).

\bibitem{Heino2003_PRE}
J.~Heino, S.~Arridge, J.~Sikora, and E.~Somersalo.
\newblock Anisotropic effects in highly scattering media.
\newblock {\em Phys. Rev. E} \textbf{68} 031908, (2003).

\bibitem{Kienle2007_PRL}
A.~Kienle.
\newblock Anisotropic light diffusion: An oxymoron?
\newblock {\em Phys. Rev. Lett.} \textbf{98} 218104, (2007).

\bibitem{Kienle2013_PhysMedBiol}
A.~Kienle, F.~Foschum, and A.~Hohmann.
\newblock Light propagation in structural anisotropic media in the steady-state
  and time domains.
\newblock {\em Phys. Med. Biol.} \textbf{58} 6205--6223, (2013).

\bibitem{Welch2010_Book}
A.~J. Welch and M.~J.~C. van Gemert (eds.).
\newblock {\em Optical-Thermal Response of Laser-Irradiated Tissue, second ed.}
\newblock Springer, (2010).

\bibitem{Alerstam2008b_JBO}
E.~Alerstam, T.~Svensson, and S.~Andersson-Engels.
\newblock Parallel computing with graphics processing units for high-speed
  Monte Carlo simulation of photon migration.
\newblock {\em J. Biomed. Opt.} \textbf{13} 060504, (2008).

\bibitem{Alerstam2010_BiomedOptExpress}
E.~Alerstam, W.~C.~Y.~Lo, T.~D.~Han, J.~Rose, S.~Andersson-Engels, and L.~Lilge.
\newblock Next-generation acceleration and code optimization for light
  transport in turbid media using GPUs.
\newblock {\em Biomed. Opt. Express} \textbf{1} 658--675, (2010).

\bibitem{Henyey1941_AstrophysJ}
L.~G. Henyey and J.~L. Greenstein.
\newblock Diffuse radiation in the galaxy.
\newblock {\em Astrophys. J.} \textbf{93} 70--83, (1941).

\bibitem{Wiersma1999_PRL}
D.~S. Wiersma, A.~Muzzi, M.~Colocci, and R.~Righini.
\newblock Time-resolved anisotropic multiple light scattering in nematic liquid
  crystals.
\newblock {\em Phys. Rev. Lett.} \textbf{83} 4321--4324, (1999).

\bibitem{Alerstam2011_PhDThesis}
E.~Alerstam.
\newblock {\em Optical spectroscopy of turbid media: time-domain measurements
  and accelerated Monte Carlo modelling}.
\newblock Ph.D. thesis, Lund University, (2011).

\bibitem{Dudko2005_BioPhysJ}
O.~K.~Dudko and G.~H.~Weiss.
\newblock Estimation of anisotropic optical parameters of tissue in a slab
  geometry.
\newblock {\em Biophys. J.} \textbf{88} 3205--3211, (2005).

\bibitem{Keijzer1988_ApplOpt}
M.~Keijzer, W.~M.~Star, and P.~R.~M. Storchi.
\newblock Optical diffusion in layered media.
\newblock {\em Appl. Opt.} \textbf{27} 1820--1824, (1988).

\bibitem{Zhu1991_PRA}
J.~X. Zhu, D.~J. Pine, and D.~A. Weitz.
\newblock Internal reflection of diffusive light in random media.
\newblock {\em Phys. Rev. A} \textbf{44} 3948--3959, (1991).

\bibitem{Wiersma2000_PRE}
D.S. Wiersma, A.~Muzzi, M.~Colocci, and R.~Righini.
\newblock Time-resolved experiments on light diffusion in anisotropic random
  media.
\newblock {\em Phys. Rev. E} \textbf{62} 6681--6687, (2000).

\bibitem{Vlahos2008_OrderandChaos}
L.~Vlahos, H.~Isliker, Y.~Kominis, and K.~Hizanidis.
\newblock Normal and anomalous diffusion: A tutorial.
\newblock In T.~Bountis, editor, {\em Order and Chaos}, Volume~10. Patras
  University Press, (2008).
\newblock (arXiv:0805.0419).

\bibitem{Contini1997_ApplOpt}
D.~Contini, F.~Martelli, and G.~Zaccanti.
\newblock Photon migration through a turbid slab described by a model based on
  diffusion approximation: I. theory.
\newblock {\em Appl. Opt.} \textbf{36} 4587--4599,  (1997).

\end{thebibliography}
\end{document}